\documentclass[preprint,aps,prc,showpacs,nofootinbib]{revtex4}
\usepackage{epsfig}
\usepackage{graphicx,amsmath}

\newcommand{\Li}{\mbox{Li}_2}

\newcommand{\dd}{\mbox{d}}
\newcommand\ba{\begin{eqnarray}}
\newcommand\ea{\end{eqnarray}}

\newcommand{\be}{\begin{equation}}
\newcommand{\ee}{\end{equation}}
\def\Li#1#2{{\mathrm{Li}}_{#1}\left(#2\right)}

\begin{document}
\title{Charge asymmetry for electron (positron)-proton elastic scattering}

\author{E. A. Kuraev, V. V. Bytev}
\affiliation{\it JINR-BLTP, 141980 Dubna, Moscow region, Russian
Federation}

\date{\today}

\begin{abstract}
Charge asymmetry in electron (positron) scattering arises from the interference of the Born amplitude and the box-type amplitude corresponding to two virtual photons exchange. It can be extracted from electron proton and positron
proton scattering experiments, in the same kinematical conditions. Considering the virtual photon Compton scattering tensor, which contributes to the box-type amplitude,
we separate proton and inelastic contributions in the intermediate state and parametrize the proton form-factors as the sum of a pure QED term and a strong interaction term. Arguments, based on analyticity, are given in favor of cancellation of contributions from proton strong interaction form factors and  of inelastic intermediate states in the box type amplitudes.  In frame of this model,  and assuming a  dipole character of form-factors, numerical estimations are given for moderately high energies.

\end{abstract}

\maketitle
\section{Introduction}

Recently, a lot of attention was devoted to $2\gamma$ exchange amplitude both in scattering and annihilation channels \cite{Re04,china,twogamma}, in connection with the experimental data on electromagnetic proton form factors (FFs) \cite{Jo00}.

Extraction of box type (two photon exchange amplitude (TPE)) contribution to elastic
electron-proton scattering amplitude
is one of long standing problems of experimental physics. It can be obtained from electron proton and positron proton scattering at the same kinematical conditions. A similar information about
TPE amplitude in the annihilation channel can be obtained from the measurement of the forward-backward asymmetry in  proton-antiproton production in  electron-positron annihilation (and the reversal process).

The theoretical description of TPE amplitude is strongly model dependent. Two reasons should be mentioned: the experimental knowledge of nucleon FFs  is restricted in a small kinematical region and the precision of the data is often insufficient, and the contribution of the intermediate hadronic states can be only calculated with large uncertainty.

A general approximation for proton electromagnetic form-factors follows the  dipole approximation:
\be
G_E(q^2)=\displaystyle\frac{G_M(q^2)}{\mu} =G_D(Q^2)=(1+Q^2/0.71\mbox{~GeV}^2)^{-2},Q^2=-q^2=-t,
\ee
where $\mu$ is the anomalous magnetic moment of proton.
However, recent experiments \cite{Jo00} showed a deviation of the proton electric FF from this prescription, when measured following the recoil polarization method, which is more precise than the traditional Rosenbluth separation. Such deviation was tentatively explained, advocating the presence of a two photon contribution.

The motivation of this paper is to perform the calculation of charge odd correlation
\begin{gather}
A^{odd}=\frac{\dd\sigma^{e^-p}-\dd\sigma^{e^+p}}{2d\sigma_B^{ep}},
\label{eq:asym}
\end{gather}
in the process of electron-proton scattering in frame of an analytical model (AM), free from uncertainties connected with inelastic hadronic state in intermediate state of the TPE amplitude. In frame of this model it is possible to show that the effects due to  strong interaction FFs and those due to the inelastic intermediate states almost completely compensate each other, within an accuracy discussed below.

Our paper is organized as follows. In part II,III we consider the charge-odd
contribution of triangle and box-type diagram. In part IV we describe the procedure of numerical integration. In In part V we present the results of numerical integration for asymmetries
and in the Conclusions we estimate the accuracy of the obtained results.
The appendices contain the tables of four-fold integrals and some details of calculations.

\section{Analytical Model formulation}
In the analysis of the TPE amplitude we consider the electromagnetic interactions in the lowest order of perturbation theory. Hadron electromagnetic FFs are functions of one kinematical variable, $Q^2$ and the static value of the Dirac FF of the proton (for $Q^2$=0) is unity due to  QED origin. Therefore we parametrize the proton FFs in the form
\be
F_1(q^2)=1+F_{1s}(q^2),\quad  F_2(q^2)=F_{2s}(q^2),\quad F_{1s}(0)=0,\quad F_{2s}(0)=\mu.
\label{eq:eq3}
\ee
Let us discuss now the arguments in favor of a cancellation of the terms of order of $F_s^2$ with the contribution of the inelastic hadronic intermediate states, in TPE amplitude.

The TPE amplitude contains the virtual photon Compton scattering tensor.
It can be splitted in two terms, when only strong
interaction contributions to Compton amplitude are taken into account. One term
(the elastic term) is the generalization of the Born term with the strong-interaction FFs at the vertexes of the interaction of the virtual photons with the hadron. We suppose that the hadron before and after the interaction with the photons remains unchanged. The second term (inelastic) corresponds to inelastic channels formed by pions and
nucleons or similar hadronic states which can be excited in the intermediate state, between the vertexes of the virtual photon interaction with the hadron.

For this aim let us present the loop momentum integration element in the form
\begin{gather}
\dd^4k=\frac{1}{2s}\dd^2k_\bot \dd s_1 \dd s_2,
\end{gather}
where $s=2pp_1$ is the total energy, $s_1=(k-p_1)^2$, and $s_2=(k+p)^2$ are the invariant mass squares of the upper (electronic) part of TPE Feynman diagram and the lower (hadronic) ones, $\dd^2k_\bot$ represents the integration on the
components of the loop momentum $k_\bot p_1=k_\bot p=0$ which are transversal  to the initial electron $p_1$ and proton $p$ momenta.

\begin{figure}
\begin{center}
\includegraphics[width=12cm]{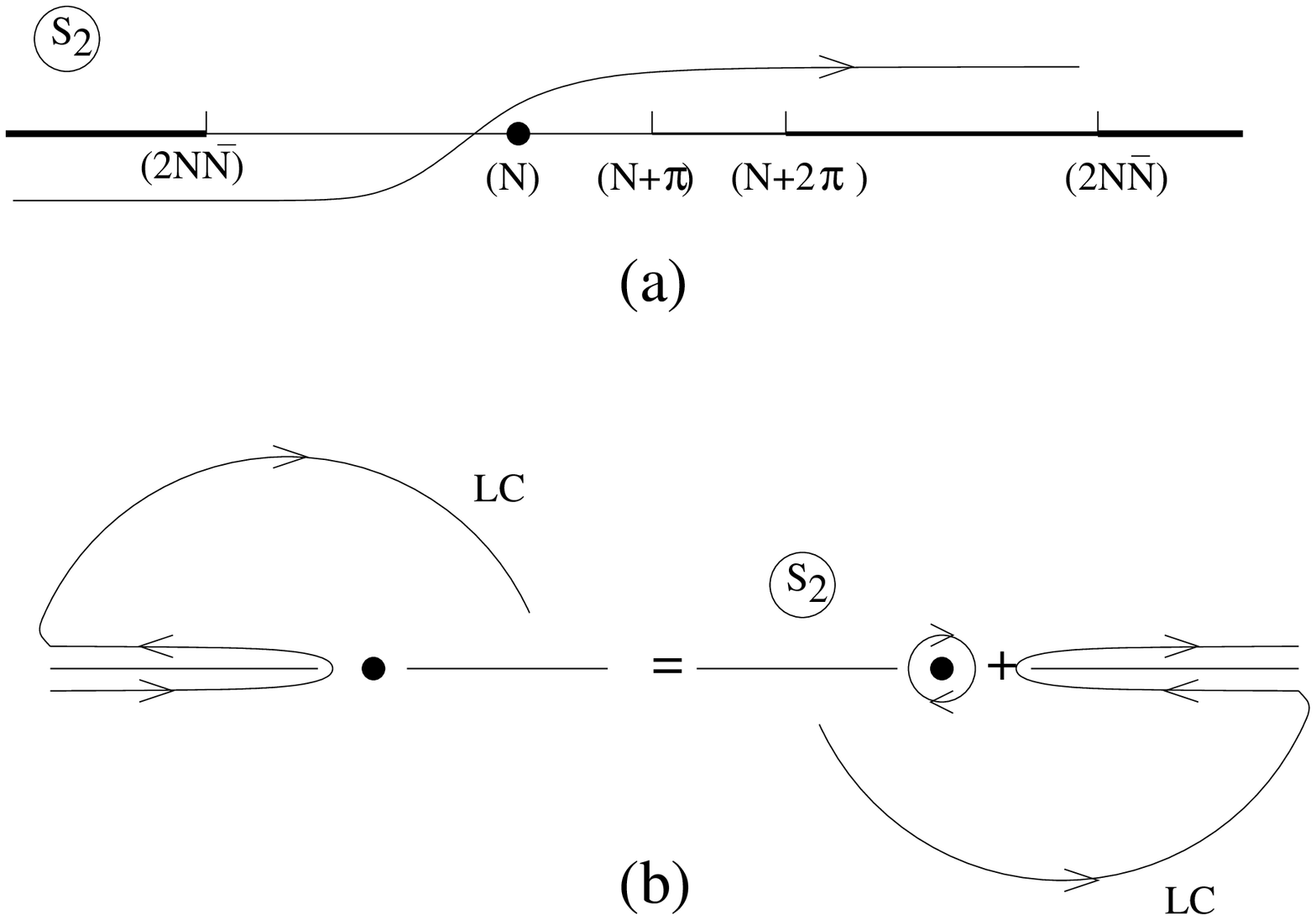}
\caption{\label{Fig:fig1}Integration contour.}
\end{center}
\end{figure}

We can consider the upper and the lower block tensors to be both gauge invariant (the factor $1/2$ is introduced to avoid  double counting and two Feynman diagrams are included in each block). Therefore, the TPE amplitude can be written in the form (we omit the factors corresponding to fermion spinors):
\ba
A=\frac{1}{2!}\frac{(4\pi\alpha)^2(2\pi i)^2}{(2\pi)^4} \int\frac{\dd^2k_\bot ds_2}{(k)(\bar{k})}
L_{\mu\nu}H^{\mu\nu},
\ea
where $L_{\mu\nu}=\gamma_\nu(\hat{p}_1-\hat{k})\gamma_\mu$, and $H_{\mu\nu}$ is the Compton tensor of proton. The integration contour is drawn
in Fig.\ref{Fig:fig1}a: it starts from $-\infty-i0$ and follows to $+\infty+i0$,
so that it belongs to the physical sheets of $s$ and $u$
channels \cite{SUMR}.

In the physical sheet, the Compton
amplitude has a pole, corresponding to a single proton state in the intermediate state and two cuts: the
right one, corresponding to the inelastic states in $s_2$-channel and the left one, starting at
$s_2<-9M^2$ ($M$ is the proton mass). Closing the integration contour to the left and to the right side
(see Fig. \ref{Fig:fig1}a) it was shown (see \cite{SUMR}) that the following relation holds
\ba
A_{left}=A_{elastic}+A_{inelastic}.
\ea
%
\begin{center}
\begin{figure}
\includegraphics[width=15cm]{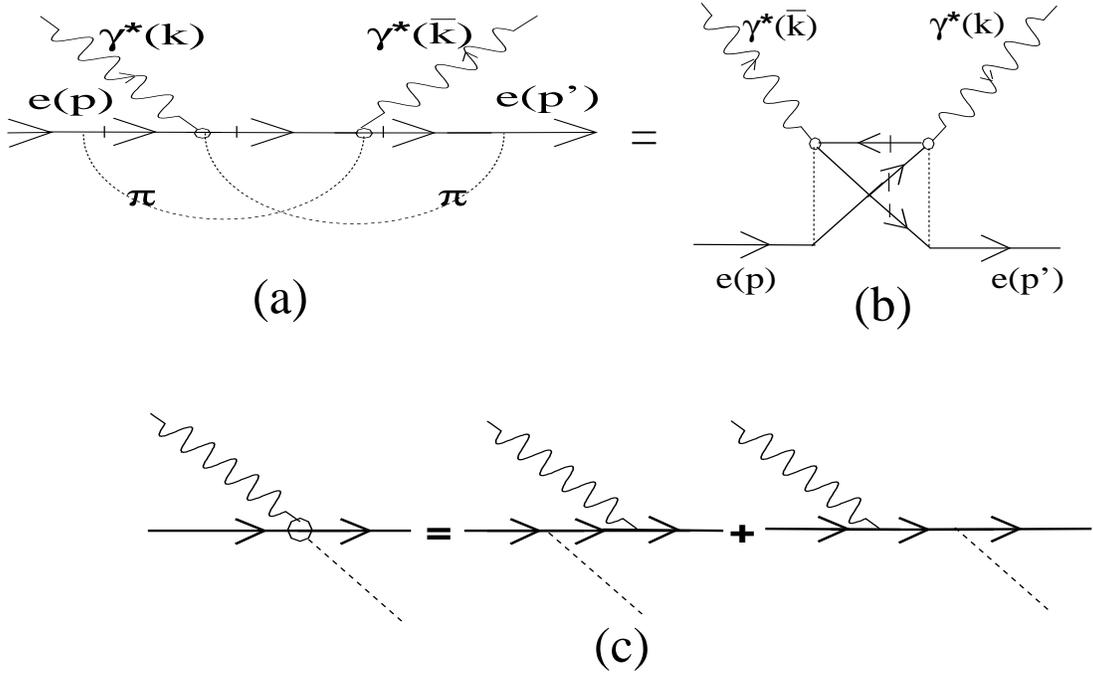}
\caption{Compton scattering Feynman diagram illustrating the left cut (a) which is equivalent to the $u$-channel discontinuity of the Feynman amplitude (b); The gauge invariant set of Compton subdiagrams is shown in (c).}
\label{Fig:fig2}
\end{figure}
\end{center}

Comparing with the sum rule obtained in \cite{SUMR}, here the Born amplitude contribution is omitted, as well as only strong interaction effects
are considered here.
Omitting the  left cut contribution $A_{left}$, (our estimate shows that it can be included in
$10$\% error bar \cite{SUMR}), the effects of strong interaction contributions to the hadron FFs compensate the strong interactions contributions arising from the inelastic channels.

As an example, the Feynman amplitude at the origin of the left cut
in the $s_2$ plane of virtual Compton scattering which contributes to the TPE blocks, is drawn on Fig. \ref{Fig:fig2}a, underlying the proton propagators which corresponds to real 3-proton u-channel intermediate state.

Its physical meaning is the interference of amplitudes of proton-
antiproton pair production in virtual photon-proton collisions due to
Fermi statistics (see Fig. \ref{Fig:fig2}b). A rough estimate of the ratio of contributions of typical right hand cut $R_{right}$ to the left hand cut $R_{left}$ is the ratio of cross sections of pion photoproduction to nucleon-antinucleon photoproduction cross section on proton:
$R_{left}/R_{right}\sim (2Mm_\pi)/(10M^2)\le 10$\%.

This argument was shown to be exact in frame of QED \cite{SUMR}, where a
rather specific kinematics was considered: forward scattering amplitude at
high energies. The application to the case of nonforward TPE amplitude requires
a more rigorous proof, which is outside the purpose of this paper. Here we  suggest to consider our approach as a model, the validity of which should be experimentally verified.

For example, experiments measuring charge-odd observables in $ep$ scattering  will be critical for verification of the validity of our model.

Proton FFs enter in the box amplitude in a form which can be schematically written as
\ba
\int \frac{d^4 k}{i\pi^2 (e)(p)} \frac{1+F_s(k^2)}{(k)}\,\,\frac{1+F_s(\bar{k}^2)}{(\bar{k})},
\ea
$$
\quad
(k)=k^2-\lambda^2, \quad (\bar{k})=(k-q)^2-\lambda^2,
\quad
(e)=(k-p_1)^2-m_e^2, \quad (p)=(k+p)^2-M^2,
$$
where we extract the QED part and do not distinguish Dirac and Pauli form
factors. This expression can be rearranged as
\ba
\frac{1+F_s(k^2)}{(k)}\,\,\frac{1+F_s(\bar{k}^2)}{(\bar{k})}&=&\frac{F_s(k^2)F_s(\bar{k}^2)}{(k)(\bar{k})}
+\frac{1}{(\bar{k})}\biggl[\frac{F_s(k^2)}{(k)}-\frac{F_s(q^2)}{q^2}\biggr]
\nonumber \\
&+&\frac{1}{(k)}\biggl[\frac{F_s(\bar{k}^2)}{(\bar{k})}
-\frac{F_s(q^2)}{q^2}\biggr]
+\frac{F_s(q^2)}{q^2}\biggl[\frac{1}{(k)}+\frac{1}{(\bar{k})}\biggr]+\frac{1}{(k)(\bar{k})}
\label{eq:4a}
\\
&=&A_{Int}+A_{FR}+A_{TR}+A_{Box}.
\label{eq:4}
\ea
According to our model assumption the first term $A_{Int}$ in the right hand side (r.h.s.) of (\ref{eq:4}) is compensated by the
inelastic intermediate hadron state contribution and so it will be omitted below. Next two terms in (\ref{eq:4a}) do not contain infrared (IR) singularities but contain the ultraviolet (UV) ones. The fourth term in (\ref{eq:4a}) suffer from both IR and UV divergences, the last one suffers only from the IR divergences. Contributions of the last two terms can be calculated analytically as well as it does not contain FFs uncertainties at the loop momentum integration. The explicit results for them are given below.

Keeping in mind the UV convergence of the initial amplitude, we extract
the UV cut-off $\Lambda$ depending contribution containing $\ln\frac{\Lambda^2}{M^2}$ from the fourth term in (\ref{eq:4a}) and add it to the contribution of the 2 and 3 terms providing their UV convergence.
This procedure denoted in equation (\ref{eq:4}). Part of calculations concerning pure QED contribution (the last term in this equation) was performed in our paper \cite{china}.

The cross-section of elastic ep scattering
\begin{gather}
e(p_1)+p(p)\to e(p_1')+p(p')
\end{gather}
in Born approximation in laboratory frame ($p=(M,0,0,0)$) has a form:
\begin{gather}
\frac{\dd\sigma_B}{\dd\Omega}=\frac{\sigma_M\sigma_{red}}{\varepsilon(1+\tau)},
\quad
\sigma_M=\frac{\alpha^2\cos^2\frac{\theta}{2}}{4E^2\sin^4\frac{\theta}{2}}\frac{1}{\rho},
\quad
\rho=1+\frac{2E}{M}\sin^2\frac{\theta}{2},
\quad \tau=\frac{Q^2}{4M^2},
\\ \nonumber
t=\frac{s(1-\rho)}{\rho},~ Q^2=-q^2=-t=2p_1p_1',\quad s=2ME,\quad u=-\frac{s}{\rho}=-2p p_1',
\\ \nonumber
s+t+u=0,\quad \varepsilon^{-1}=1+2(1+\tau)\tan^2\frac{\theta}{2},
\end{gather}
with
\begin{gather}
\sigma_{red}=\tau G_M^2+\varepsilon G_E^2,\quad
 G_M=F_1+F_2, \quad G_E=F_1-\tau F_2.
\end{gather}
Here $\theta$- electron scattering angle and $F_1=1+F_{1s}$,
$F_2=F_{2s}$- Dirac and Pauli proton's FFs.

Keeping in mind the representation in the form of Eq. (\ref{eq:4}),
it will be convenient to use another (equivalent) form of Born cross section:
\begin{gather}
\frac{\dd\sigma_B}{\dd\Omega}
=\frac{\dd\sigma_{Bt}}{\dd\Omega}+\frac{\dd\sigma_{Bbox}}{\dd\Omega}
=\frac{\alpha^2}{M^2\rho^2t^2}(B_t+B_{box}),
\\ \nonumber
B_t=\frac{1}{2}(F_1^2-F_1)(2tM^2+s^2+u^2)+t^2F_1F_2-\tau F_2^2(tM^2+s u)
-\frac{1}{2}F_2 t^2,
\nonumber \\
 B_{box}=\frac{1}{2}F_1(2tM^2+s^2+u^2)+\frac{1}{2}t^2F_2.
\label{eq:brn}
\end{gather}

The IR divergence from virtual photon emission contribution is, as usually, canceled when summing with contribution from emission of soft real photons:
\begin{gather}
\frac{\dd\sigma^{soft}}{\dd\Omega}=
\biggl[\frac{\dd\sigma_{Bt}}{\dd\Omega}
+\frac{\dd\sigma_{Bbox}}{\dd\Omega}
\biggr]\delta_{soft}^{odd}
=\frac{\dd\sigma_{Bt}^{soft}}{\dd\Omega}
+\frac{\dd\sigma_{Bbox}^{soft}}{\dd\Omega}.
\end{gather}

 The quantity
$\delta_{soft}^{odd}$
was considered in \cite{china}, \cite{maximon}:
\ba
\delta_{soft}^{odd}&=&-2\left . \frac{4\pi\alpha}{16\pi^3}\int\frac{d^3k}{\omega}(\frac{p_1'}{p_1'k}-\frac{p_1}{p_1k})
(\frac{p'}{p'k}-\frac{p}{pk})\right |_{S_0,\omega\le \Delta E}
\nonumber \\
&=&\frac{2\alpha}{\pi}\biggl[2\ln\frac{1}{\rho}\ln\frac{2\rho\Delta E}{\lambda}
+\ln x\ln\rho+\Li{2}{1-\frac{1}{\rho x}}-\Li{2}{1-\frac{\rho}{x}}\biggr],
\nonumber \\
x&=&\frac{\sqrt{1+\tau}+\sqrt{\tau}}{\sqrt{1+\tau}-\sqrt{\tau}},
\ea
with $\lambda,\Delta E$- soft photon mass and it's maximal energy in
laboratory frame.

Note that $\delta_{soft}^{odd}$ does not have a definite symmetry under
substitution $ p\leftrightarrow-p';s\leftrightarrow u $ due to the specific
definition of soft photon in the laboratory frame.

The virtual contribution to the cross section can be splitted in three
terms (according Eq. (\ref{eq:4})):
\begin{gather}
\frac{\dd \sigma_v}{\dd\Omega}=
\frac{\dd \sigma_{vt}}{\dd\Omega}+\frac{\dd \sigma_{vb}}{\dd\Omega}
+\frac{\dd \sigma_F}{\dd\Omega}
=\frac{\alpha^3}{2\pi t^2M^2\rho^2}(a_t+a_b+a_f).
\end{gather}
The first term appears from the contribution of triangle-type diagram
and can be put in the form:
\begin{gather}
a_t=(1-P(s\leftrightarrow u))\int\frac{\dd^4 k}{i\pi^2}\frac{1}{(k)}
\biggl[
\frac{S_eS_T}{(e)(p)}+\frac{S_{\bar{e}}S_{\bar{T}}}{(\bar{e})(\bar{p})}
\biggr],
\end{gather}
with
\begin{gather}
(\bar{e})=(k+p_1')^2-m_e^2,~(\bar{p})=(p'-k)^2-M^2,
\end{gather}
where $P(s\leftrightarrow u)f(s,u)=f(u,s)$ is the substitution operation and
\ba
S_e&=&\frac{1}{4}Tr\hat{p}_1'\gamma_\mu(\hat{p}_1-\hat{k})\gamma_\nu\hat{p}_1\gamma_\eta,
\nonumber \\
S_{\bar{e}}&=&\frac{1}{4}Tr\hat{p}_1'\gamma_\mu(\hat{p}_1'+\hat{k})\gamma_\nu\hat{p}_1\gamma_\eta,
\nonumber \\
S_T&=&\frac{1}{4}Tr R((\Gamma_{\mu}(q)-\gamma_{\mu})(\hat{p}+\hat{k}+M)\gamma_{\nu},\nonumber \\
S_{\bar T}&=&\frac{1}{4}Tr R\gamma_{\mu}(\hat{p}'-\hat{k}+M)(\Gamma_{\nu}(q)-\gamma_{\nu}),\nonumber \\
\Gamma_{\mu}(q)&=&\gamma_\mu F_{1}(q^2)+\frac{1}{4M}F_{2}(q^2)(\hat{q}\gamma_\mu-
\gamma_\mu \hat{q});~R=(\hat{p}+M)\Gamma_{\eta}(-q)(\hat{p}'+M).
\ea

The box type contribution is parameterized as:
\begin{gather}
a_b=(1-P(s\leftrightarrow u))\int\frac{\dd^4 k}{i\pi^2}\frac{ t S_e Z_p}{(k)(\bar{k})(p)(e)},
\\ \nonumber
Z_p=\frac{1}{4}Tr R\gamma_\mu(\hat{p}+\hat{k}+M)\gamma_\nu.
\end{gather}
Some of the necessary integrals have been previously calculated in \cite{KM76a}. However, for completeness, they are all given in Appendix A.

The finite part contribution (2nd and 3rd terms in rhs of Eq. (\ref{eq:4}))
are parametrized as:
\be
a_f=(1-P(s \leftrightarrow u))\int\frac{\dd^4 k}{i\pi^2}
\frac{t}{(\bar k)}\biggl[ \frac{S_eS_F}{(e)(p)}
+\frac{S_{\bar{e}}S_{\bar{F}}}{(\bar e)(\bar p)}
\biggr]
=
A_f(s,u)-A_f(u,s),
\ee
and
\ba
S_{F}&=&\frac{1}{4}Tr R\gamma_\mu(\hat{p}+\hat{k}+M)\Phi_\nu; \nonumber\\
S_{\bar{F}}&=&\frac{1}{4}Tr R \Phi_\mu(\hat{p}'-\hat{k}+M)\gamma_\nu,
\nonumber \\
\Phi_\mu&=&\left [\frac{1}{(k)} (F_1(k^2)-1)-\frac{1}{q^2}
(F_1(q^2)-1)\right ]\gamma_\mu -
\frac{1}{4M} \left [\frac{[\hat{q}\gamma_\mu]}{q^2}F_2(q^2)-
\frac{[\hat{k},\gamma_\mu ]}{(k)}F_2(k^2)\right ].
\label{eq:eq23}
\ea
In the integration over the four-vector $k$ in $a_f$ and $a_t$
enters the UV cut-off parameter $|k|^2<\Lambda^2$.
To obtain  $A_{FR}$ and $A_{TR}$ in Eq. (\ref{eq:4}) the terms
containing $\ln(\Lambda^2/M^2)$ are singled out from the contribution of the fourth term
of rhs (\ref{eq:4a}) and add it to the contribution of 2nd and 3rd term in
(\ref{eq:4a}). So we can put in (17) $a_t+a_b+a_f=a_{tr}+a_b+a_{fr}$,
explicitly eliminating the cut-off parameter dependence.

\section{Virtual and soft photon emission contributions of triangle type diagram}

The interference of Born amplitude with the part of TPE arising from the fourth
term from the r.h.s. of Eq. (\ref{eq:4}) with the corresponding part of soft photon emission leads to
\ba
\displaystyle\frac{\dd\sigma_{T}}{\dd\Omega}
&=&\displaystyle\frac{\dd\sigma_{vt}}{\dd\Omega}
+\displaystyle\frac{\dd\sigma_{Bt}^{soft}}{\dd\Omega}
\nonumber\\
&=&\displaystyle\frac{2\alpha^3}{M^2\rho^2t^2\pi}
\left \{
2B_t\left [
\ln\displaystyle\frac{1}{\rho}\ln\frac{2\Delta E}{M}-\ln^2\rho
+\displaystyle\frac{1}{2}\ln\rho\ln x+\displaystyle\frac{1}{2}\Li{2}{1-\frac{1}{\rho x}}\right .\right .
\nonumber \\
&&\left .\left .-\frac{1}{2}\Li{2}{1-\frac{\rho}{ x}}
\right ]+\frac{1}{2}D_{tsv}
\right \}+\frac{3\alpha^3(u-s)}{2M^2\rho^2t\pi}(F_2+F_1)(1-F_1)\ln\frac{\Lambda^2}{M^2},
\label{eq:tr}
\ea
with
\ba
D_{tsv}&=&
B_t\biggl[\ln^2\frac{s}{M^2}-\ln^2\frac{-u}{M^2}
-2\Li{2}{1+\frac{M^2}{s}}
+2\Li{2}{1+\frac{M^2}{u}}-\pi^2 \biggr]
 \nonumber \\
&&
+[1-P(s\leftrightarrow u)]\left (A(s,t)\ln\frac{s}{M^2}+B(s,t)\right ),
\ea
and
$$A(s,t)=a_1F_1(1-F_1)+a_2F_2+a_3F_2^2+a_4F_1F_2,$$
$$B(s,t)=b_1F_1(1-F_1)+b_2F_2+b_3F_2^2+b_4F_1F_2,$$
with
\ba
a_1&=&\frac{M^4(M^2+t)}{2(M^2+s)^2}(3M^2+4s)+\frac{1}{4}[-6M^4+4sM^2+2tM^2+6u^2],
 \nonumber \\
a_2&=&\frac{tM^4(7M^2+9s)}{4(M^2+s)^2}-\frac{t}{4}(5u+7M^2), \nonumber \\
a_3&=&-\frac{M^2t}{16(M^2+s)^2}(2M^4+5st+3sM^2+5tM^2)+\frac{t}{16M^2}(2M^4+M^2(u-2t)-8su), \nonumber \\
a_4&=&-\frac{M^2t}{8(M^2+s)^2}(8M^4+2st+23sM^2+2tM^2)-\frac{t}{8}(-18M^2+18t+13s), \nonumber \\
b_1&=&-\frac{s}{4}(7t+2M^2)-\frac{M^4(M^2+t)}{2(M^2+s)}, \quad
b_2=-\frac{7}{4}st-\frac{tM^4}{2(M^2+s)}; \nonumber \\
b_3&=&\frac{st}{16}+\frac{tM^4}{16(M^2+s)},   \quad
b_4=\frac{15}{8}st+\frac{5tM^4}{8(M^2+s)}.
\ea
Note that function $D_{tsb}$ in r.h.s. of Eq.(\ref{eq:tr})
changes the sign at substitution $s\leftrightarrow u$,
in particular
\begin{gather}
\rho=\frac{s}{-u}\to \frac{1}{\rho},\quad
Re\biggl[\ln^2\frac{-s-i0}{M^2}\biggr]=\ln^2\frac{s}{M^2}-\pi^2\to\ln^2\frac{-u}{M^2};
~\ln \frac{s} {M^2}\to \ln \frac{-u} {M^2} .
\end{gather}
The last term in (24), which contains the cut-off parameter is necessary in order to remove the
$\Lambda$-dependence of $a_f$.

\section{Virtual and soft photon emission contributions of QED box type diagram}

The box-type contribution (last term in (\ref{eq:4})) with corresponding part
of soft photon emission is given by (the list of necessary loop momentum integrals
and details of the calculation are given in Appendix A):
\begin{eqnarray}
\displaystyle\frac{\dd\sigma_{B}}{\dd \Omega}
&=&\displaystyle\frac{\dd\sigma_{vb}}{\dd \Omega}
+\displaystyle\frac{\dd\sigma_{Bbox}^{soft}}{\dd \Omega_{e}}\nonumber \\
&=&
\displaystyle\frac{2\alpha^{3}}{\pi M^{2}\rho^{2}t^2}
\left \{
B_{box}\left [
\ln\frac{1}{\rho}\ln\displaystyle\frac{(2\Delta E)^2}{4\tau M^2}-2\ln^2\rho
+\ln\rho\ln x \right .\right .
\nonumber \\
&&+ \left .\left .\Li{2}{1-\displaystyle\frac{1}{\rho x}}-\Li{2}{1-\displaystyle\frac{\rho}{ x}}
\right ]
-\frac{t^2}{2}
[1-P(s\leftrightarrow u)]
(d_1F_1-d_2F_2) \right \},
 \label{eq:eq23a}
\end{eqnarray}
with
\begin{eqnarray}
d_1&=&\frac{s}{t}
\left [
\displaystyle\frac{1}{2}
\ln^{2}( 4\tau   )
-\frac{\tau}{1+\tau}\ln ( 4\tau   )
+M^{2}{F}_{Q}\left ( 6\tau +2 -\frac{2\tau^{2}}{1+\tau}\right )
\right .
\nonumber \\
&-&\left .\left .\ln^{2}\frac{s}{-t} +\pi^{2}
+2\Li{2}{ 1+\frac{M^{2}}{s}}\right ]\right .
\nonumber \\
&-&\frac{(1-2\tau)}{4\tau}
\left [
2\ln \rho \ln ( 4\tau )-\ln^{2}\frac{s}{M^{2}}   +\pi^{2}+
2\Li{2} {1+\frac{M^{2}}{s}}\right ]
 + \left(2M^{2}-\frac{su}{t} \right ) \frac{\ln\frac{s}{M^{2}}}{s+M^{2}};
\nonumber \\
d_2&=&\frac{s}{2(1+\tau)}\left [F_Q- \frac{1}{2 M^{2}}\ln ( 4\tau   )  \right ]
- \frac{M^{2}}{M^{2}+s} \ln\frac{s}{M^2}-  \ln \rho \ln ( 4\tau )
+\frac{1}{2} \left ( \ln^{2}\frac{s}{M^{2}}-\pi^{2}\right )\nonumber \\
&&
- \Li{2}{ 1+\frac{M^{2}}{s}},
\label{eq:eq24}
\end{eqnarray}
and $F_Q$ given in (\ref{eq:Fq}).

\section{Infrared singularities free contributions of box amplitudes}

In this section we use the following ansatz for nucleon FFs
\be
F_1(q^2)=F_2(q^2)/\mu=\left (\frac{Q_0^2}{q^2-Q_0^2}\right )^2,
\label{eq:eq30}
\ee
setting the parameter $Q_0^2$  to 0.71 GeV$^2$. This form permits us to carry on analytical calculations.
Note that this presctiption differs from the ones given above, Eq. (1).
Let us rewrite the expressions Eq. (\ref{eq:eq23}) for $\Phi_{\mu}$ as follows:
\be
\Phi_{\mu}=\Phi_1\gamma_{\mu}- \frac{[\hat{q}\gamma_\mu]}{4M}F_2(t)\Phi_2-
\mu\frac{[\hat{q}-\hat{k},\gamma_{\mu} ]}{4M}\Phi_3,
\label{eq:eqa1}
\ee
with (here we use dipole approximation for FFs, Eq. \ref{eq:eq30}):
\ba
\Phi_1&=&\frac{ (q^2-k^2)}{(Q_k)^2}[A(Q_k)+B],~
A=\frac{2Q^2_0-q^2}{(Q_q)^2}=\frac{1-F_1(t)}{t},~B=\frac{Q_0^2}{(Q_q)},
\nonumber \\
\Phi_2&=&\frac{ ( k^2-q^2)}{(k)t(Q_k)^2 }
\left [(Q_k)^2 +q^2(Q_k)+q^2(Q_q) \right ],
\nonumber \\
\Phi_3&=&\frac{(Q_0^2)^2}{(k)(Q_k)^2 },~(Q_q)=q^2-Q_0^2,~(Q_k)=k^2-Q_0^2.
\label{eq:eqphi}
\ea
When integrating on the loop momenta, the $\Phi_1$ and $\Phi_2$ terms give origin to
UV divergences contributions containing $L_{\Lambda}=\ln\frac{\Lambda ^2}{M^2}$ in the form:
\be
A_f^{\Lambda }(s,u)\simeq \int dx y dy\left [\frac{1-F_1(t)}{t}\left (\ln\frac{\Lambda ^2}{d_0}-\frac{3}{2}\right ) G(s,t) +
\frac{F_2(t)}{4Mt}\left (\ln\frac{\Lambda ^2}{{\cal D}_0}-\frac{3}{2}\right ) F(s,t)\right ],
\label{eq:eqa2}
\ee
with $D_0,d_0$ given in Appendix B, and
\ba
G(s,t)&=& -\displaystyle\frac{1}{4} Tr\hat{ p}_1'\gamma_\mu\gamma_\sigma \gamma_\nu \hat{p}_1 \gamma_\eta \cdot
\displaystyle\frac{1}{4} Tr  R \gamma_\mu\gamma_\sigma\gamma_\nu
=-F_2(q^2)(2t^2-6st) -F_1(q^2) (2u^2+8s^2+10 tM^2);
\nonumber\\
F(s,t)&=& -\displaystyle\frac{1}{4} Tr \hat{ p}_1'\gamma_\mu\gamma_\sigma \gamma_\nu \hat{ p}_1 \gamma_\eta \cdot
\displaystyle\frac{1}{4} Tr  R \gamma_\mu\gamma_\sigma {[\hat{q},\gamma_\nu ]}
=-F_2(q^2)\left (8Mt^2+ \frac{8stu}{M}\right ) -16t^2M F_1(q^2).
\label{eq:eqa3}
\ea
The $\Phi_2$ and $\Phi_3$ terms contain IR divergences. However, in both regions where IR divergences are present i.e., $k\to 0$, $k\to q$, the sum of the  contributions  $\Phi_2+\Phi_3$ converge.
As for UV contributions, we note that the quantity:
\be
a_{fr}\simeq A_f^\Lambda(s,u)-A_f^\Lambda(u,s)-3(u-s)(F_2+F_1)(F_1-1)L_{\Lambda}, ~L_{\Lambda}=\ln \frac{\Lambda^2}{M^2},
\ee
is finite at the limit $\Lambda\to\infty$. It results in the replacement $\Lambda^2=M^2$, $L_{\Lambda}=0.$
The explicit expression for $a_{fr}$ in terms of three-fold integrals is given in Appendix B.
We note that UV divergences associated with $\Phi_2$ are canceled due to the symmetry $F(s,t)=F(u,t)$.

The relevant contribution to the differential cross section can be written as:
\be
\frac{d\sigma_{F}}{d\Omega}= \frac{d\sigma_B}{d\Omega}\frac{\alpha}{\pi}D_f,~  D_f=\frac{a_{fr}}{2(B_t+B_{box})}.
\label{eq:eq41}
\ee
\section{Results and Discussion}

The differential cross section with two photon exchange and the relevant soft photon emission is given by:
\be
\displaystyle\frac{d\sigma}{d\Omega}= \displaystyle\frac{d\sigma_{Born}}{d\Omega}
+\displaystyle\frac{d\sigma_T}{d\Omega}
+\displaystyle\frac{d\sigma_B}{d\Omega}+\displaystyle\frac{d\sigma_F}{d\Omega}.
\label{eq:eq41a}
\ee
Our result for charge asymmetry (\ref{eq:asym}) has the form:
\begin{gather}
A^{odd}=\frac{\alpha}{\pi}\biggl[
2\ln\frac{1}{\rho}\ln\frac{2\Delta E}{M}+{\cal D}(\frac{E}{M},\theta)
\biggr],~{\cal D}(\frac{E}{M},\theta)=D_{tbs}+D_f;
\end{gather}
where $D_{tbs}$ denotes the analytical part :
\begin{eqnarray}
D_{tbs}&=&\ln x\ln\rho-2\ln^2\rho+\Li{2}{1-\frac{1}{\rho x}}-\Li{2}{1-\frac{\rho}{ x}}
\nonumber \\
&&
+\frac{1}{B_t+B_{box}}\left \{
B_t \left [
\ln^2\frac{s}{M^2}-\ln^2\frac{-u}{M^2}-\pi^2
-2\Li{2}{1+\frac{M^2}{s}}
+2\Li{2}{1+\frac{M^2}{u}}\right ]\right .
\nonumber \\
&&
+ B_{box}\ln \rho\ln(4\tau)
\nonumber \\
&&
+ \left .\left[1-P(s\leftrightarrow u)\right ] \left [A(s,t) \ln
\frac{s}{M^2}+B(s,t)-t^2(d_1F_1-d_2F_2)
\right ]\right \}.
\label{eq:eq42}
\end{eqnarray}
We note that charge asymmetry $A^{odd}$ is finite in zero electron mass limit, but contains the soft photon emission parameter $\Delta E/M$.

The function $D_{tbs}$, which can be calculated analytically and the function $D_{f}$, Eq. (\ref{eq:eq42}), for which a numerical integration has been performed, have been calculated for several values of angles and energies. Both contributions are larger (in absolute value) at large $E/M$.

The ansatz (\ref{eq:eq3}) which reflects the possibility to separate the QED and the strong interactions contribution to FFs and that assumes $F_{QED}=1$ is useful as it allows to perform, at least partly, analytical calculations. However, it can not be considered exact. A better parametrization of QED and strong interaction contributions to FFs is :
\be
F_1(Q^2)=F_{1Q}(x)+F_{1s}(x),~ F_{1Q}(x)=\frac{1+x^2}{(1+x)^4},~ F_{1s}(x)=\frac{2x}{(1+x)^4},~~x=\frac{-q^2}{Q_0^2}=\frac{Q^2(~\mbox{GeV})}{0.71}.
\label{eq:eqff}
\ee

Therefore, in order to find the global correction due to two-photon exchange,
one should weight the individual contributions $D_{tbs}$ and $D_{f}$,
by the ratio of strong and EM FFs. This can be done by multiplying $D_{tbs}$ by
the factor $F_{1Q}(x)$. The total contribution is therefore:
$$D\to\tilde {\cal D}=F_{1Q}(x)(D_{tbs}+D_{f}).$$

Tabulated values of the numerical results are presnted in Table \ref{Table:table1}.
One can see that the two photon contribution to
the asymmetry is of the order of  percent, keeping in mind the multiplicative
factor $\alpha/\pi$. The behavior is smooth in the considered kinematical ranges.
Singularities are expected for $\theta=0$ $\epsilon=1$, due to symmetry properties
of the $2\gamma$ exchange \cite{Re04}.

Taking into account the factor $\alpha/\pi$, the corrections do not exceed 1\%, in the considered kinematical range.

\section{Conclusions}

Charge asymmetry in electron proton elastic scattering contains essential information on the contribution of $2\gamma$ exchange to the reaction amplitude. This amplitude can shed light on Compton scattering of virtual photons on proton. It contains a part corresponding to proton intermediate state, which carries the information on proton FFs. Another term corresponds to excited nucleon states and inelastic states such as $N\pi$, $N2\pi$, $N\bar N N$. Their theoretical investigation is strongly model dependent.

Our main assumption about the compensation of pure strong interaction induced contributions to FFs and inelastic channels allows us to avoid additional uncertainty connected with inelastic channels.

Our choice of photon form factors, (see f. (\ref{eq:eq30})) in nn physical one.
The physical case corresponds to
\begin{gather}
G_E=\frac{1}{\mu}G_M=\frac{Q_0^4}{(Q_0^2-q^2)^2}
\end{gather}
The aim of the choice is the simplification of the analytical calculation.

The assumption about the possibility to omit $A_{int}$ in (9) was proved to hold in frame of QED, and for $ep$ elastic scattering for the kinematics of almost forward scattering. The application to large angle scattering requires, in principle, a rigorous proof.

The parameterization of the $NN^*\gamma^*$ vertex is also approximated, since one of the nucleon is off mass shell. Nevertheless, they can be estimated to be small for $Q^2 <$  5 GeV$^2$ and included in the sources of theoretical uncertainties. The reliability of our assumption can be estimated from the ratio of cross sections of pion and nucleon antinucleon pair photoproduction. The uncertainty of our results does not exceed 10\%.

Another interesting question is if such compensation is present for the annihilation channel. The measurement of charge asymmetry is, in this case, associated with polar angle odd contribution to the differential cross section.

Similar effects of charge and angular asymmetries  can also be due to Z-boson exchange, but such contribution is small for moderate-high energy colliders. The ratio of corresponding contributions  can be evaluated as: $\sim (\pi g_Vg_A s)/(\alpha M_Z^2)< 5 \cdot 10^{-3}$ for $s<10$ GeV$^2$ ($g_V$ ($g_A$) is the vector (axial) coupling constant of the Z boson to fermion).

The analytical calculation of $2\gamma$ amplitude with FFs encounters mathematical difficulties. In Ref. \cite{Bo06}
the results for the box amplitude with arbitrary FFs was investigated. Similar attempt was done by A. Ilichev \cite{Il06}. These works use different approaches to include FFs, however the numerical results are close to ours, and show that two photon contribution can not be responsible for the discrepancy in recent FFs measurements.

Other works \cite{Ko05} devote much attention to the excited intermediate states as $\Delta$ and $N^*$ resonances, introducing additional uncertainties. In our approach, excited states should not be included, as they correspond to poles in the second physical sheet.

Our numerical results show that charge-odd correlations are of the order of percent, in the kinematical region considered here. Such value is expected to be larger at larger $q^2$ values and could be measured in very precise experiments, at present facilities.

\section{Acknowledgments}
One of us (E.A.K.) acknowledges the kind hospitality of Saclay,
where part of this work was done, and especially to prof. Egle Tomasi-Gustafsson
for critical remarks and help. This work was partly supported by
grant ÌÊ-2952.2006.2. The authors are grateful to C. C. Adamu\v s\v c\'in for technical help.

\section{Appendix A. List of necessary integrals.}
\label{app:A}
We give here a list of scalar, vector and tensor type loop momentum integrals with three
denominators $(k),(e),(p)$:
$$
\int\frac{\dd^4k}{i\pi^2}\frac{1;k_\mu;k_\mu k_\nu}{(k)(e)(p)}=Z_1;\,\,Z_2 p_{1\mu}+Z_3 p_\mu;
$$
\ba
Z_4 g_{\mu\nu}+Z_5p_{1\mu}p_{1\nu}+
Z_6p_\mu p_\nu+Z_7(p_{1\mu}p_\nu+p_{1\nu}p_\mu).
\ea
Standard procedure of joining the denominators leads to integrals of the form
\ba
\int\limits_0^1 \dd x\int\limits_0^12y
\dd y\int\frac{1;k_\mu;k_\mu k_\nu}{[(k-yp_x)^2-y^2p_x^2-\lambda^2(1-y)]^3},
\ea
with $p_x=xp_1-(1-x)p,p_x^2=m^2x^2+M^2(1-x)^2+s_1x(1-x)$, $s_1=-s-i0$.
Further integration on loop momentum (see (37))
with $\Lambda$-is the UV cut-off parameter, leads to:
\ba
Z_1&=&\frac{1}{2s}\biggl[L_s^2-\frac{1}{2}L^2-2Li_s+\ln\frac{M^2}{\lambda^2}(2L_s+L)\biggr],\quad
L=\ln\frac{M^2}{m^2},\quad L_s=\ln\frac{s}{M^2}-i\pi,
\nonumber \\
Li_s&=&Li_2(1+\frac{M^2}{s})-i\pi\ln(1+\frac{M^2}{s}),
\nonumber \\
Z_2&=&\frac{1}{s}\biggl[L+(1+\frac{M^2}{s+M^2})L_s\biggr],\quad Z_3=-\frac{1}{s+M^2}L_s,\quad
\nonumber \\
Z_4&=&\frac{1}{4}L_\Lambda+\frac{3}{8}-\frac{s}{4(s+M^2)}L_s,\nonumber \\
Z_5&=&\frac{1}{2s}\biggl[-1-\frac{M^2}{s+M^2}+L_s(1-\frac{M^4}{(s+M^2)^2})+L\biggr],\quad\nonumber \\
Z_6&=&\frac{1}{2}\frac{1}{s+M^2}\biggl[\frac{s}{s+M^2}L_s-1\biggr],
\nonumber \\
Z_7&=&-\frac{1}{2}\frac{1}{s+M^2}\biggl[1+\frac{M^2}{s+M^2}L_s\biggr].\nonumber
\ea
Integrals with denominators $(k)(\bar{e})(\bar{p})$ can be obtained from ones given above by replacements
$p_1\to-p_1'$, $p\to -p'$ with the same coefficients.
Integrals with denominators $(k)(\bar{e})(p)$ can be obtained from ones given above by replacements
$p_1\to-p_1'$, $p\to p,$ and coefficients, which can be obtained from the ones mentioned above
by replacement $s\to u$.

Box type integrals defined as
\begin{gather}
Y_1;I_\mu;I_{\mu\nu}=\int\frac{d^4k}{i\pi^2}\frac{1;k_\mu;k_\mu k_\nu}{(0)(q)(e)(p)};\quad
I_\mu=Y_2 \Delta_\mu+Y_3 P_\mu,
\nonumber \\
I_{\mu\nu}=Y_4 g_{\mu\nu}+Y_5\Delta_\mu\Delta_\nu
+Y_6P_\mu P_\nu+Y_7(P_\mu\Delta_\nu+P_\nu\Delta_\mu)
+Y_8Q_\mu Q_\nu,
\end{gather}
with
\begin{gather}
Q=\frac{p_1-p_1'}{2},\quad P=\frac{p_1+p_1'}{2},\quad \Delta=\frac{p+p'}{2}.
\end{gather}
Explicit expressions for $Y_k$ are:
\begin{eqnarray}
Y_1&=&\frac{2}{st}\ln\frac{-s-i0}{Mm}\ln\frac{-t}{\lambda^2},
\nonumber \\
Y_2&=&-\frac{1}{2d}\biggl[
-\frac{\tau}{2}(F+F_Q)-P^2(F+F_\Delta)
\biggr],
\quad
Y_3=\frac{1}{2d}\biggl[
-\frac{\tau}{2}(F+F_\Delta)-\Delta^2(F+F_Q)
\biggr],
\nonumber \\
Y_4&=&\frac{1}{\tau}\biggl[
-\frac{\tau}{2}(F-G+H_p+H_\Delta+H_Q)+H_\Delta(-\frac{\tau}{2}-P^2)
-H_Q(-\frac{\tau}{2}-\Delta^2)
\nonumber \\
&&+2Q^2Y_3(P^2+\tau)+P^2G_\Delta-\Delta^2G_Q+2Q^2\Delta^2Y_2
\biggr],
\nonumber  \\
Y_5&=&\frac{1}{\tau d}
\biggl[
P^2\frac{\tau}{2}H+2(P^2)^2(H_\Delta-2Q^2Y_3-G_\Delta)
+(\frac{\tau^2}{4}-2P^2\Delta^2)(H_Q+2Q^2Y_2-G_Q)
\biggr],
\nonumber \\
Y_6&=&\frac{1}{\tau d}\biggl[
-\frac{\tau}{2}\Delta^2(G-F-H_p-3H_\Delta+6Q^2Y_3)
+(P^2\Delta^2+\frac{\tau^2}{4})(H_\Delta-2Q^2Y_3-G_\Delta)
\nonumber \\
&&-\Delta^2(H_Q+2Q^2Y_2-G_Q)
\biggr],
\nonumber \\
Y_7&=&-\frac{1}{\tau d}\biggl[
\frac{\tau^2}{4}H-\Delta^2\frac{\tau}{2}(H_Q+2Q^2Y_2-G_Q)
+2P^2\frac{\tau}{2}(H_\Delta-2Q^2Y_3-G_\Delta)
\biggr],
\nonumber \\
Y_8&=&-\frac{1}{Q^2\tau}
\biggl[
-\tau(H_\Delta-2Q^2Y_3+H_P+\frac{1}{2}F-\frac{1}{2}G)+\Delta^2(H_Q-2Q^2Y_3-G_Q)
\nonumber \\
&&-P^2(H_\Delta-2Q^2Y_3-G_\Delta)
\biggr],
\nonumber \\
\tau&=&2P\Delta,\quad d=P^2\Delta^2-\frac{\tau^2}{4},\quad H=F-G+H_P+3H_\Delta+6P^2Y_3.
\end{eqnarray}
The quantities entering here are:
\ba
M^2F_Q&=&-\frac{1}{4\sqrt{\tau(1+\tau)}}\left [
\pi^2+\ln(4\tau)\ln x+\Li{2}{-2\sqrt{\tau x}}-\Li{2}{\frac{2\sqrt{\tau}}{\sqrt{x}}}\right ],
\nonumber \\
F_\Delta&=&-\frac{1}{t}\biggl [\frac{1}{2}\ln^2\frac{-t}{m^2}+\frac{\pi^2}{6}\biggr ],\quad
G_Q=-\frac{1}{4M^2(1+\tau)}\biggl [-tF_Q-2\ln\frac{-t}{M^2}\biggr ],
\nonumber \\
F&=&\frac{1}{2s}\biggl[2\ln\frac{-s-i0}{Mm}\ln\frac{-t}{M^2}-\ln^2(\frac{-s-i0}{M^2})+
2\ln^2\frac{M}{m}+2\Li{2}{1+\frac{M^2}{s}}\biggr],
\nonumber \\
H_Q&=&\frac{1}{s+M^2}\ln\frac{-s-i0}{M^2}.
\label{eq:Fq}
\ea
The explicit expression for QED box-type Born amplitude (see Eq. (21)) is
$ a_b(s,u)=A_b(s,u)-A_b(u,s)$ with
\begin{eqnarray}
A_b(s,u)&=&(F_1P_q-F_2Q_q) F_Q + (F_1P_{\Delta}-F_2Q_{\Delta})F_\Delta + (F_1P_G-F_2Q_G)G_Q+\nonumber \\
&&
(F_1P_H-F_2Q_H) H_Q + (F_1P_F-F_2Q_F)F + (F_1P_Y-F_2Q_Y)Y_1,
\nonumber
\end{eqnarray}
with
\begin{eqnarray}
P_q&=& \frac{1}{8}\left [M^2(s-u)+s(3s-u)\right ];~ Q_q=\frac{st}{8};~P_\Delta= \frac{M^2t}{4} + \frac{3s^2}{8};
\nonumber \\
Q_\Delta&= &-\frac{t^2}{8};~ P_G= -\frac{M^2t}{4} - \frac{tu}{8};~Q_G= \frac{M^2t}{4} - \frac{st}{8};
\nonumber \\
P_H&=& \frac{M^2t}{2}-\frac{su}{4};~Q_H=- \frac{tM^2}{4};~P_F= \frac{sM^2}{2}+\frac{s(s-u)}{4};~
\nonumber \\
Q_F&=&- \frac{st}{4};~P_Y=-\frac{M^2ts}{2} - \frac{s(s^2+u^2)}{4};~ Q_Y= \frac{st^2}{4}.
\label{eq:Fq1}
\end{eqnarray}
\section{Appendix B. }
\label{app:B}
The contributions from the uncrossed box-type Feynman amplitude, $A_f$, can be written as the sum of terms associated to $\Phi_1$, $\Phi_2$, $\Phi_3$:

$$A_f=A_{f1} -\displaystyle\frac{1}{4M}\left (F_2(t) A_{f2}+(Q_0^2)^2 t\mu A_{f3}\right ),~ a_{fr}=(1-P(s\leftrightarrow u))A_f.$$
We follow the procedure described in Appendix C.

The $\Phi_2$ and $\Phi_1$ terms are calculated using the relation:
\be
\frac{k^2-q^2}{(k-q)^2}=1-2q_{\alpha}\frac{(q-k)_{\alpha}}{(\bar k)},
\label{eq:eqb4}
\ee
in order to eliminate explicitly the divergence at $k\to q$. For $A_{f1}$ we have:
\ba
A_{f1}&=&t \int_0^1 dx y dy \left \{G(s,t)\left [A\left (\ln\frac {M^2}{d_0}-\frac{3}{2}
\right )+\frac{B\bar{y}}{d_0}\right ]
-2N(yp_x) \left (\frac{A}{d_0} +\frac{B\bar{y}}{d_0^2}\right )\right \}+\nonumber\\
&&
t \int_0^1 dx y dy z^2dz \left \{
{\cal S} \left (\frac{A}{D_1}+\frac{B\bar{z}}{D_1^2}\right )
-2N(zP_0) \left ( \frac {A}{D_1^2}+\frac {2B\bar{z}}{D_1^3}\right )
t(2\bar{z}+yz)\right \}
\ea
with
$${\cal S} =tG(s,t)(2\bar{z}+yz)+2(z\bar{y}-\bar{z})[2t(s^2+tM^2)F_1-2st^2F_2],$$
and
$$N(b)=\frac{1}{4} Tr \hat p_1'\gamma_\mu(\hat{p}_1-\hat{b})\gamma_\nu \hat{p}_1\gamma_\eta\cdot
\frac{1}{4} Tr R \gamma_\mu(\hat{p}+\hat{b}+M)\gamma_\nu, $$
and $A,B$ given above (Eq. (\ref{eq:eqphi})).
The expression for  $A_{f2}$ is:
\ba
A_{f2}&=&\int dx dy y z^2 dz \left \{
F(s,t)\left [ \left (\ln\frac {\Lambda^2}{D_0}-\frac{3}{2}\right )-q^2\bar{y}L_1+q^2(Q_q)\cdot \bar{y}^2L_2\right ] \right . \nonumber\\
&&\left . + \left [-\frac{1}{D_0}+q^2\bar yL_2-2q^2(Q_q)\bar y^2L_3\right ][M(yp_x)+\bar M(yp'_x)]
\right \}
\nonumber\\
&&
-2\int  dx dy y z^2 dz \left \{\left [-\frac{1}{2a}+\frac {q^2\bar z}{2}{\cal J}_2-q^2(Q_q)\bar z^2
{\cal J}_3\right ]\right .
q^{\alpha}\left [F_{\alpha}(zP_0)+\bar F_{\alpha}(zP_0')\right ]\nonumber\\
&&
\left .+ \frac{t[2\bar z+zy]}{2}
\left [\frac{1}{a^2}-2\bar z q^2{\cal J}_3+6q^2(Q_q)\bar z^2{\cal J}_4
\right ][M(zP_0)+\bar M(z P_0')] \right \},
\ea
with ${\cal J}_i,L_i$ given in Appendix C and

$$ M(b)=\frac{1}{4} Tr\hat p_1'\gamma_\mu(\hat{p}_1-\hat{b})\gamma_\nu \hat{p}_1\gamma_\eta\cdot
\frac{1}{4} Tr R \gamma_\mu(\hat{p}+\hat{b}+M)[\hat{q},\gamma_\nu ],$$
$$\bar M(b)= \frac{1}{4} Tr \hat{p}_1'\gamma_\mu(\hat{p}_1'+\hat{b})\gamma_\nu \hat{p}_1\gamma_\eta\cdot
\frac{1}{4} Tr R [\hat{q},\gamma_\mu ](\hat{p}'-\hat{b}+M)\gamma_\nu.$$
\ba
F_{\alpha}(b)&=&\frac {1}{4}Tr \hat p_1'\gamma_{\mu}\gamma_{\alpha}\gamma_\nu \hat p_1 \gamma_\eta \cdot
\frac {1}{4}Tr R \gamma_{\mu}(\hat p +\hat b+M)[\hat q,\gamma_\nu] -\nonumber\\
&&\frac {1}{4}Tr \hat p_1'\gamma_{\mu} (\hat p_1-\hat b)\gamma_\nu\hat  p_1\gamma_\eta
  \cdot
\frac {1}{4}Tr R \gamma_{\mu} \gamma_{\alpha}[\hat q,\gamma_\nu] + ( q - b)^{\alpha}F(s,t);
\ea
and similar expression for $\bar F_{\alpha}$.

The form of $A_{f3}$ is:

\be
A_{f3}= \int_0^1dx ydy(z\bar z)^2dz
\left \{-[H(zP_0)+\bar H(zP_0')]{\cal J}_3+
6[G(zP_0)+\bar G(zP_0')]{\cal J}_4\right \}
\label{eq:AppB1}
\ee
with
\ba
G(k)&=&\frac{1}{4}Tr\hat{p}_1'\gamma_\mu(\hat{p}_1-\hat{k})\gamma_\nu\hat{p}_1\gamma_\eta\cdot
\frac{1}{4}Tr  R \gamma_\mu(\hat{p}+\hat{k}+M) [\hat{q}-\hat{k},\gamma_\nu];\nonumber\\
\bar{G}(k)&=&\frac{1}{4}Tr\hat{p}_1'\gamma_\mu(\hat{p}_1'+\hat{k})\gamma_\nu\hat{p}_1\gamma_\eta\cdot
\frac{1}{4}Tr  R [\hat{q}-\hat{k},\gamma_\mu] (\hat{p}'-\hat{k}+M)\gamma_\nu ;\\
H(b)&=&\frac{1}{4}Tr \hat{p}_1'\gamma_\mu\gamma_\sigma\gamma_\nu \hat p_1 \gamma_\eta
\left \{ -\frac{1}{4}Tr R \gamma_\mu\gamma_\sigma [\hat{q}-\hat{b},\gamma_\nu]+
\frac{1}{4}Tr R \gamma_\mu(\hat{p}+\hat{b}+M)[\gamma_\sigma,\gamma_\nu]\right\}\nonumber\\
&&-\frac{1}{4}Tr \hat{p}_1'\gamma_\mu(\hat{p_1}-\hat{b}) \gamma_\nu \hat{p}_1\gamma_\eta\cdot
\frac{1}{4}Tr  R \gamma_\mu\gamma_\sigma[\gamma_\sigma,\gamma_\nu].\nonumber\\
\bar H(b)&=&\frac{1}{4}Tr \hat{p}_1'\gamma_\mu\gamma_\sigma\gamma_\nu \hat p_1 \gamma_\eta
\left \{ -\frac{1}{4}Tr R [\gamma_\sigma\gamma_\mu] (\hat{p}'-\hat{b}+M)\gamma_\nu-
\frac{1}{4}Tr R [\hat{q}- \hat{b},\gamma_\nu]\gamma_\sigma\gamma_\nu \right\}\nonumber\\
&&+\frac{1}{4}Tr \hat p_1'\gamma_\mu(\hat{p_1}'+\hat{b}) \gamma_\nu
\hat{p}_1\gamma_\eta\cdot\frac{1}{4}Tr  R [\gamma_\sigma\gamma_\mu]\gamma_\sigma\gamma_\nu.\nonumber\\
\label{eq:eqb3}
\ea
Infrared singularities (divergences of $A_{f2}$, $A_{f3}$
 at $y\to 0$, $z\to 1$ are mutually compensated in the sum $A_f$.
\section{Appendix C. }
\label{app:C}
Let us describe the procedure employed for compacting the denominators and for the loop momentum integration.
Taking ${\cal D}(z)=a z+b\bar z$, $\bar z=1-z$, the Feynman prescription for compacting the denominators leads to:
\ba
\frac {1}{ab}&=&\int_0^1 \frac {dz}{{\cal D}(z)^2};~
\frac {1}{a^2b}=\int_0^1 \frac {2zdz}{{\cal D}(z)^3};~
\frac {1}{a^2b^2}=\int_0^1 \frac {6z\bar zdz}{{\cal D}(z)^4};~\nonumber \\
\frac {1}{a^3b}&=&\int_0^1 \frac {3z^2 dz}{{\cal D}(z)^4};~
\frac {1}{a^3b^2}=\int_0^1 \frac {12z^2 \bar zdz}{{\cal D}(z)^5};~
\frac {1}{a^3b^3}=\int_0^1 \frac {30 (z\bar z)^2 dz}{{\cal D}(z)^6}.
\label{eq:eqf}
\ea
Applying (\ref{eq:eqf}) to the set of denominators:
$$(e)=k^2-2kp_1,~(\bar e)=k^2+2kp_1',~(p)=k^2+2kp;$$
$$~(\bar p)=k^2-2p'k,~(k)=K^2, (\bar k)=(q-k)^2, (Q_k)=k^2-Q_0^2,$$
one obtains:
\ba
\frac {1}{(epk)}&=\int_0^1 \displaystyle\frac {2dx ydy}{[k-yp_x)^2-{\cal D}_0]^3};
&\displaystyle\frac {1}{(ep\bar k)}=\int_0^1
\displaystyle\frac {2dx ydy}{[(k-P_0)^2-{\cal D}_0]^3};\nonumber \\
\displaystyle\frac {1}{(\bar e\bar p\bar k)}&=\int_0^1 \displaystyle\frac {2dx ydy}{[(k-P_0')^2-{\cal D}_0]^3};
&\displaystyle\frac {1}{(e p Q_k)}=\int_0^1\displaystyle\frac {2dx ydy}{[(k-yp_x)^2-d_0]^3};
\label{eq:eqf1}
\ea
with ${\cal D}_0=y^2p_x^2$; $d_0=y^2p_x^2+\bar y Q_0^2$, $P_0=yp_x+\bar y q$,
$P_0'=y p_x'+\bar y q$, $p_x=xp_1-\bar xp$, $p_x'=\bar xp'-xp_1'$. The following relation holds: $P_0^2=P_0'^2=y^2p_x^2+\bar y q^2$,
$p_x^2=p_x'^2=\bar x^2M^2-sx\bar x$.

We do not develop further (explicitly) the denominators which contain $(\bar e) (\bar p)$, because the corresponding results can be obtained by those depending on $(e)(p)$ under replacement $p_x\to p_{x'}$.

In a similar way one obtains:
\ba
&&\displaystyle\frac{1}{(epQ_k^2)}=\int_0^1 \displaystyle\frac{6y\bar ydx dy}{[(k-yp_x)^2- d_0]^4};
\displaystyle\frac{1}{(kepQ_k)}=\int_0^1 \displaystyle\frac{6y\bar ydx dydt}{[(k-yp_x)^2- d_t]^4};
~d_t={\cal D}_0+t\bar y Q_0^2;\nonumber\\
&&\displaystyle \frac{1}{(kepQ_k^2)}=\int_0^1 \displaystyle \frac{24 dx t dt y\bar y^2 dy}{[(k-yp_x)^2- d_t]^5};
\displaystyle\frac{1}{(ep\bar k Q_k)}=\int_0^1
\displaystyle\frac{6y dy dx z^2dz}{[(k-zP_0)^2- {\cal D}_1]^4};~{\cal D}_1=a+\bar z Q_0^2;\nonumber\\
&&\displaystyle \frac{1}{(epk\bar k)}=\int_0^1 \displaystyle \frac{6ydydxz^2dz}{[(k-zp_0)^2-a]^4};
a=z^2P_0^2-z\bar yq^2;
\nonumber\\
&&\displaystyle \frac{1}{(epk\bar k Q_k)}=\int_0^1
\displaystyle \frac{24ydy z^2 \bar zdz dt dx}{[(k-zP_0)^2-{\cal D}_2]^5};~
{\cal D}_2=a+\bar zQ_0^2t;\nonumber\\
&&\displaystyle \frac{1}{(epk\bar k Q_k^2)}=\int_0^1 \displaystyle
\frac{120 ydy dx tdt (z \bar z)^2dz}{[(k-zP_0)^2- {\cal D}_2]^6}.
\ea
After replacing the loop momenta by $k-b \to \kappa$, a symmetrical integration on $\kappa$ is performed. For polynomials in $N(k)$ of order in  $k$ $\le 3$  one finds:
$N(\kappa +b)=\frac {1}{4} \kappa^2(b)+N(b)$. Applying Wick rotation leads to:
\ba
&\int\displaystyle \frac {\kappa^2 d\kappa} {(\kappa^2-d)^3}=\ln\frac{\Lambda^2}{d} -\frac{3}{2};~
&\int\displaystyle \frac {d\kappa} {(\kappa^2-d)^3}=-\displaystyle\frac{1}{2d};\nonumber\\
&\int\displaystyle \frac {\kappa^2 d\kappa}{(\kappa^2-d)^4}=-\frac{1}{3d};
&\int\displaystyle \frac { d\kappa}{(\kappa^2-d)^4}=\displaystyle\frac{1}{6d^2};\nonumber\\
&\int\displaystyle \frac {\kappa^2 d\kappa}{(\kappa^2-d)^5}=\displaystyle\frac{1}{12d^2};
&\int\displaystyle \frac {d\kappa}{(\kappa^2-d)^5}=-\displaystyle\frac{1}{12d^3};\nonumber\\
&\int\displaystyle \frac {\kappa^2 d\kappa}{(\kappa^2-d)^6}=-\displaystyle\frac{1}{30d^3};
&\int\displaystyle \frac { d\kappa}{(\kappa^2-d)^6}=\displaystyle\frac{1}{20d^4};
\label{eq:eqc5}
\ea
with
$$d\kappa= \frac {d^4\kappa}{i\pi^2}=
\kappa^2_e d\kappa_e^2,~\kappa^2 =-\kappa^2_e,~
\kappa^2_e =\kappa^2_0+\kappa^2_1+\kappa^2_2+\kappa^2_3, $$
and $\kappa_e$ - euclidean four vector.
Note that the integration of the Feynman parameter $t$ can be provided explicitly, as it enters only in phase volume:
\ba
L_1&=&\int_0^1\displaystyle\frac{dt}{d_t}= \displaystyle \frac{1}{\bar y Q_0^2} \ln  \displaystyle \frac{d_0}{{\cal D}_0};
L_2=\int_0^1 \displaystyle \frac{dt}{d_t^2}= \displaystyle \frac{1}{{\cal D}_0d_0} ;
\nonumber\\
L_3&=&\int_0^1 \displaystyle\frac{t dt}{d_t^3}= \displaystyle \frac{1}{(\bar y Q_0^2)^2}
\left [\ln\displaystyle \frac{t d_0}{{\cal D}_0}- 1+\displaystyle \frac{{\cal D}_0}{d_0} \right ];
\nonumber \\
{\cal J}_2&=&\int_0^1 \displaystyle \frac{dt}{{\cal D}_2^2}=
\displaystyle \frac{a+{\cal D}_1}{2a^2{\cal D}_1^2};
\nonumber \\
{\cal J}_3&=&\int_0^1 \displaystyle \frac{tdt}{{\cal D}_2^3}= \displaystyle \frac{1}{2a{\cal D}_1^2} ;
{\cal J}_4=\int_0^1 \displaystyle \frac{tdt}{{\cal D}_2^4}= \displaystyle \frac{{\cal D}_1+ 2a}
{6a^2{\cal D}_1^3}.
\label{eq:eqb5}
\ea

\begin{center}
\begin{table}[ht]
\begin{tabular}{|c|c|c|c|c|c|}
\hline
$E/M$-$\theta$  &30 & 60 & 90 &120&150 \\
\hline\hline
    1 & 1.16  & 0.58  & -0.72 &  -1.29 & -1.81  \\
\hline
    2 & 3.71 & 1.19  &  -0.22 &  -0.96 & -1.61 \\
\hline
    3 & 4.23 & 1.54  & -0.15 & -0.97 & -1.73 \\
 \hline
    4 &  4.46 & 1.86 & 0.32 & -0.62 & -1.35 \\
 \hline
  5   & 4.52  & 2.43 & 1.12 & -0.13 & -0.99 \\
\hline
\end{tabular}
\caption[]{ Numerical values of $\tilde{\cal D}$ as a function of $E/M$ and $\theta$, with  dipole parametrization of the form factors.}
\label{Table:table1}
\end{table}
\end{center}

\end{document}